\documentclass[aps,prl,twocolumn,letterpaper,superscriptaddress]{revtex4-1}
\usepackage{amsmath}
\usepackage{dcolumn}
\usepackage{epsfig}
\usepackage{graphicx}
\usepackage{latexsym}
\usepackage{amssymb}
\usepackage{color}
\usepackage{amsfonts}
\usepackage{bm}

\newcommand{\pco}{Pr$_2$CuO$_4$}

\newcommand{\pccox}{Pr$_{1.86}$Ce$_{0.14}$CuO$_{4\pm\delta}$}
\newcommand{\pccod}{Pr$_{2-x}$Ce$_x$CuO$_{4\pm\delta}$}
\newcommand{\ncco}{Nd$_{2-x}$Ce$_x$CuO$_4$}

\begin{document}
\title{Shubnikov-de Haas quantum oscilations reveal a reconstructed Fermi surface near optimal doping in a thin film of the cuprate superconductor \pccox}

\author{Nicholas P. Breznay*}
\affiliation{Department of Physics, University of California, Berkeley, Berkeley CA 94720, USA}
\affiliation{Materials Sciences Division, Lawrence Berkeley National Laboratory, Berkeley CA 94720, USA}
\altaffiliation{nbreznay@berkeley.edu}
\author{Ian M. Hayes}
\affiliation{Department of Physics, University of California, Berkeley, Berkeley CA 94720, USA}
\affiliation{Materials Sciences Division, Lawrence Berkeley National Laboratory, Berkeley CA 94720, USA}
\author{B. J. Ramshaw}
\affiliation{NHMFL, Los Alamos National Laboratory, Los Alamos, New Mexico 87545, USA}
\author{Ross D. McDonald}
\affiliation{NHMFL, Los Alamos National Laboratory, Los Alamos, New Mexico 87545, USA}
\author{Yoshiharu Krockenberger}
\affiliation{NTT Basic Research Laboratories, NTT Corporation, 3-1 Morinosato-Wakamiya, Atsugi, Kanagawa 243-0198, Japan}
\author{Ai Ikeda}
\affiliation{NTT Basic Research Laboratories, NTT Corporation, 3-1 Morinosato-Wakamiya, Atsugi, Kanagawa 243-0198, Japan}
\author{Hiroshi Irie}
\affiliation{NTT Basic Research Laboratories, NTT Corporation, 3-1 Morinosato-Wakamiya, Atsugi, Kanagawa 243-0198, Japan}
\author{Hideki Yamamoto}
\affiliation{NTT Basic Research Laboratories, NTT Corporation, 3-1 Morinosato-Wakamiya, Atsugi, Kanagawa 243-0198, Japan}
\author{James G. Analytis}
\affiliation{Department of Physics, University of California, Berkeley, Berkeley CA 94720, USA}
\affiliation{Materials Sciences Division, Lawrence Berkeley National Laboratory, Berkeley CA 94720, USA}

\date{\today}

\begin{abstract}

We study magnetotransport properties of the electron-doped superconductor \pccod\ with $x$ = 0.14 in magnetic fields up to 92~T, and observe Shubnikov de-Haas magnetic quantum oscillations. The oscillations display a single frequency $F$=255$\pm$10~T, indicating a small Fermi pocket that is $\sim$~1\% of the two-dimensional Brillouin zone and consistent with a Fermi surface reconstructed from the large hole-like cylinder predicted for these layered materials. Despite the low nominal doping, all electronic properties including the effective mass and Hall effect are consistent with overdoped compounds. Our study demonstrates that the exceptional chemical control afforded by high quality thin films will enable Fermi surface studies deep into the overdoped cuprate phase diagram.

\end{abstract}

\maketitle

\begin{figure}[bth]
\includegraphics[width=0.9\columnwidth]{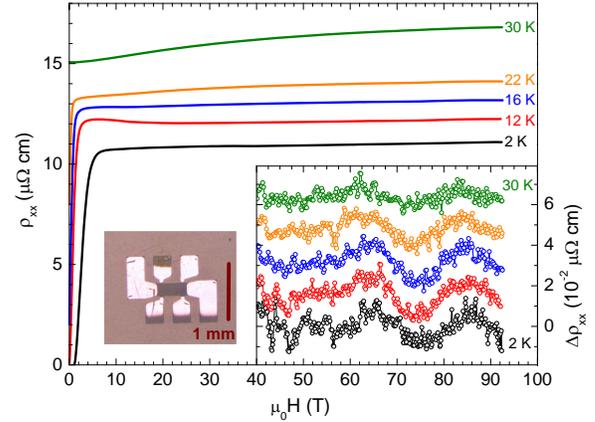}
\caption{(Color online) Low-temperature magnetoresistance of a superconducting \pccod\ thin film (pictured) measured to 92~T at temperatures between 2~K and 30~K. Inset (right scale): sample magnetoresistance with a smooth background subtracted, showing magnetic oscillations that are suppressed with increasing temperature.}
\label{f:mr}
\end{figure}

Understanding the ordering phenomena that compete or coexist with superconductivity in the cuprate superconductors remains an outstanding challenge.  Central to this effort is identification of Fermi surface (FS) topology and evolution with doping via studies of magnetic quantum oscillations (QO), led by initial observation of QO in hole-doped YBa$_2$Cu$_3$O$_{6.5}$\cite{doiron2007, leboeuf2007}. In hole-doped cuprates, QO studies have shown (1) a large cylindrical hole-like FS consistent with band theory in overdoped Tl$_2$Ba$_2$CuO$_{6+\delta}$~\cite{vignolle2008}, (2) FS reconstruction and a complex topology in underdoped YBa$_2$Cu$_3$O$_{6+\delta}$~\cite{sebastian2014}, and (3) strong enhancements in the quasiparticle effective mass~\cite{ramshaw2015}; see Refs.~\onlinecite{vignolle2011,sebastian2012} for further reviews. These observations have been interpreted as evidence for competing electronic ordered phases and the influence of quantum critical fluctuations, crossing over to Fermiology consistent with the band-theory picture. Comparable experimental studies of electron-doped cuprates are limited, and precise description of their FS remains an outstanding challenge. Here we report on the FS topology and effective mass in the electron-doped cuprate \pccod\ (PCCO) with $x=0.14$. We observe Shubnikov-de Haas QO in a PCCO thin film measured in extreme magnetic fields up to 92~T, where magnetotransport data (Fig.~\ref{f:mr}) show evidence for a small (255~T) FS pocket, a light quasiparticle effect mass $m/m_e = 0.43$, and direct determination of the orbitally averaged Fermi velocity $v_F = 2.4 \times 10^5$~m/s.

Electron-doped cuprates \textit{R}$_{2-x}$Ce$_x$CuO$_{4\pm\delta}$ (with \textit{R} = La, Nd, Pr, ...) have a square planar $T'$ structure with two CuO$_2$ layers within the body-centered tetragonal unit cell. Bandstructure calculations predict a large, hole-like FS cylinder centered at $(\pi,\pi)$ arising from the CuO$_2$ planes~\cite{massidda_electronic_1989}. (See Ref.~\onlinecite{armitage2010} for a recent review.) This cylinder (discussed below) should contain $n = 1 - x$ carriers assuming the electron dopant concentration is equal to the Ce content $x$ in \pccod. Photoemission (ARPES) experiments are consistent with a large FS in overdoped $n$-doped materials~\cite{armitage2002, matsui2007, song2012}, and show evidence for both electron- and hole-like pockets as $x$ decreases below $\sim 0.16$. However, many issues remain unresolved, including the structure of the FS and nature of its reconstruction and evolution with doping, reconciling quantum oscillation measurements with the observation of Fermi arcs~\cite{shen_nodal_2005,reber_origin_2012} in the pseudogap phase, and identifying the true nature of the competing ground state (whether arising from observed magnetic~\cite{motoyama_spin_2007,saadaoui2015} charge~\cite{dasilvaneto2015}, or predicted d-density wave (DDW)~\cite{chakravarty_hidden_2001}) order). QO have only been observed in a restricted doping range ($x$ = 0.15 - 0.17) in $n$-doped bulk crystals of \ncco\ (NCCO)~\cite{helm2009,helm2010,helm2015}, and found to be consistent with a theoretical picture of DDW-like order~\cite{eun2010j,eun2011c}. Resolving QO in Ce-doped PCCO thin films opens an exciting avenue for continuously studying FS evolution, in particular because thin films allow doping levels beyond the $x$ = 0.17 solubility limit of conventional bulk synthesis techniques.

\begin{figure}[tb]
    \includegraphics[width=1.0\columnwidth]{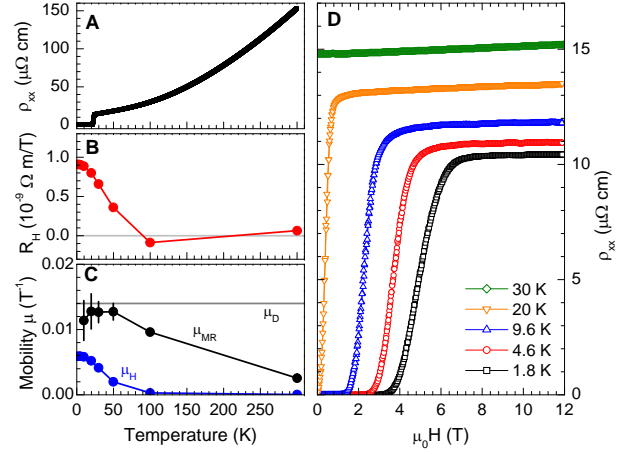}
\caption{(Color online) Resistivity and magnetotransport in a $x$ = 0.14 \pccod\ thin film. (A) $\rho$-T curve; the sample shows a superconducting transition at  $T_c$ = 22~K and a residual resistivity ratio $\equiv \rho(300K)/\rho(30K)$ = 10.2.  (B) Hall coefficient $R_H = \rho_{yx} / B$ as a function of temperature. (C) Hall ($\mu_{H}$), magnetoresistance ($\mu_{\textrm{MR}}$), and Dingle ($\mu_{D}$) mobilities calculated as described in the text. (D) $\rho$(B) magnetoresistance traces measured in DC fields at temperatures between 1.8 and 30~K.}
\label{f:rvst}
\end{figure}

We study a superconducting PCCO film (thickness 100~nm) that was grown using molecular beam epitaxy and characterized as described in detail elsewhere~\cite{krockenberger2012}. Cerium content $x = 0.14$ is controlled to within 1 \% via an in-situ quartz-crystal thickness monitor and ICP spectroscopy. Hall-bar devices with active area 200~$\times$~350 $\mu m^2$ were defined using conventional photolithography techniques, as shown in the inset of Fig.~\ref{f:mr}.  We measured in-plane resistivity $\rho_{xx}$ and Hall effect $\rho_{yx}$ using standard four-point lock-in configurations in DC magnetic fields (to 12~T); applied fields were parallel to the crystallographic $c$-axis. Figure~\ref{f:rvst} shows the temperature dependent resistivity $\rho_{xx}$, Hall coefficient $R_H \equiv \rho_{yx}/B$, and mobility $\mu$ for this \pccox\ film; while $\rho_{yx}$(B) is linear in this field range, high field measurements have shown nonlinearity and evidence for multi-band behavior~\cite{li2007}. Film parameters are summarized in Tab.~\ref{tab:params}, below. The film shows a sharp superconducting transition with $T_c$ = 22~K and a zero-temperature upper critical field $H_{c2} = 5$~T. The residual resistance ratio, defined as $\rho(300 K) / \rho(30 K)$ = 10.2, while the residual resistance is $\approx$15 $\mu\Omega$~cm. The carrier mobility estimated both from the Hall effect in a single-band picture $\mu_H = R_H / \rho_{xx}$, and parabolic component of the magnetoresistance ($\Delta R / R \sim (\mu_{\textrm{MR}} B)^2$), is $\approx 0.01$~m$^2$/V\,s; this is in good agreement with the mobility extracted from QO analyses (see below) and indicates that the strong-field limit of $\omega_c \tau \sim$~1 can be accessed in applied fields $\sim 100$~T. (Here $\omega_c = e B/m^*$ is the cyclotron frequency, and $\tau$ the quasiparticle lifetime.)

We measured the high-field magnetoresistance in pulsed magnetic fields to over 92~T at the NHMFL Pulsed Field Facility. Measurements were carried out with the sample in He-4 and He-3 liquid (2 - 4~K) or gas (5 - 30~K), and care was taken to minimize heating effects during field pulses. The data reported here are from the rising field portion of the pulsed measurements; some hysteresis was observed and is reflected in the quoted uncertainties, but the QO analyses are consistent between rising and falling field sweeps. Figure~\ref{f:mr} shows the central result of this work: \pccox\ film magnetoresistance as a function of applied magnetic fields to above 92~T, at temperatures between 2~K and 30~K. In sufficiently strong magnetic fields, the separation between quasiparticle Landau levels can be greater than their lifetime $\tau$ and thermal broadening. With changing field, the resulting oscillations of the density of states lead to Shubnikov de-Haas oscillations in the conductivity, visible in the magnetoresistance data after a smooth background has been subtracted (Fig.~\ref{f:mr}, inset) with an amplitude that increases with increasing field and decreasing temperature. The appearance and nature of these QO allow for direct observation and study of FS properties.

\begin{figure}[tb]
    \includegraphics[width=1.0\columnwidth]{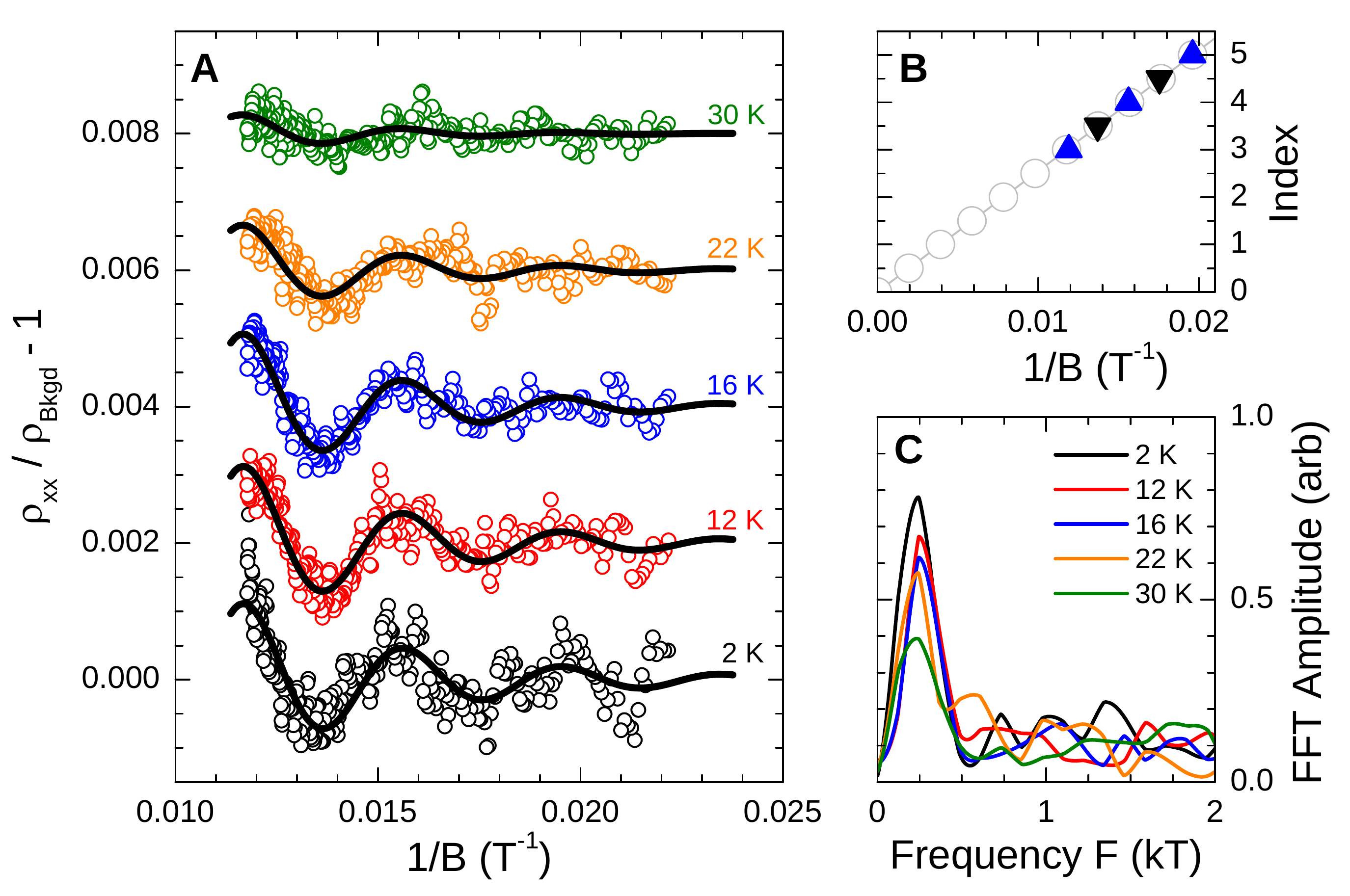}
\caption{(Color online) (A) After subtracting a polynomial background, magnetic quantum oscillations periodic in inverse field are visible below 0.02~T$^{-1}$, with period 1/255 T$^{-1}$. (Curves have been vertically offset for clarity.) (B) Location of maxima (up triangles) and minima (down triangles) of the QO traces versus inverse field. (C) FFTs of the QO data, showing a single peak near 250~T.}
\label{f:invb}
\end{figure}

To analyze the QO visible in the low-temperature magnetoresistance, we fit to and divide by 3rd-order polynomial background $\rho(B) = \rho_{\text{bkgd}}(1 + \textit{f}_{\textrm{QO}}(B))$ to reveal oscillations periodic in inverse field shown in the main panel of Fig.~\ref{f:invb}. The oscillation frequency $F = 255\pm10$~T is consistent with both Landau level indexing and by computing Fast Fourier transform (FFT) spectra for all temperatures (also plotted in Fig.~\ref{f:invb}). The frequency $F = \left( \frac{\hbar}{2 \pi e} \right) A_F$ is determined by the extremal FS cross sectional area $A_F$ perpendicular to applied field; here $A_F = 2.4 \times 10^{18}$ m$^{-2}$ or 1.0\,\% of the two-dimensional (2D) Brillouin zone area. In contrast, a band-filling picture using the nominal Ce concentration $x$ predicts a carrier concentration $p = 1 - x \approx$ 0.86 holes per Cu, or 43\% of the 2D Brillouin zone. At 2~K $R_H = 0.9 \times 10^{-9}$~$\Omega$\,m/T; assuming a single parabolic band yields a carrier density per CuO$_2$ layer of $n_{2D}$ = 4.2 $\times 10^{18}$ m$^{-2}$ or 0.66 carriers per Cu. As has been seen for near-optimal doping in both PCCO thin films~\cite{dagan2004,li2007,charpentier2010} and bulk crystals of \ncco ~\cite{helm2015}, $R_H$ changes sign with temperature, consistent with a multi-band FS and possible onset of competing order. (As $R_H$ also changes sign with $x$ near optimal doping, the simplistic analysis of the Hall effect discussed here will necessarily be incomplete.) Assuming a quasi-2D FS cylinder with parabolic dispersion, the Fermi energy can be directly calculated from $F$ yielding $E_F$ = 24~meV $\sim$ 280~K. Finally, following a model for reconstruction of the large hole-like FS with a ($\pi,\pi$) ordering wavevector~\cite{helm2009}, we estimate an energy gap $\Delta \sim $30~meV.

With a circular orbit, $A_F = \pi k_F^2$ and the orbitally averaged Fermi velocity $v_{F} = \hbar k_F / m^* = 2.4 \times 10^{5}$~m/s. Though $v_F$ is somewhat below that determined by ARPES measurements of \ncco\ $4.3 \times 10^{5}$~m/s~\cite{armitage2003}, it is several times larger than that reported recently in the hole-doped cuprate materials Bi$_2$Sr$_2$CaCu$_2$O$_{8+\delta}$ and YBa$_2$Cu$_3$O$_{6.5}$ near optimal doping, $7.7-8.4\times 10^4$~m/s~\cite{vishik_doping_2010, jaudet_dehaas_2008}. Interestingly, $v_F$ is within 10\% of the speculative ``universal'' nodal value~\cite{zhou_high_2003} observed in ARPES measurements in many hole-doped cuprate materials across a wide range of doping levels, hinting that these QOs originate from a nodal (hole-like) FS pocket.

The QO evolution with temperature and field are in excellent agreement with fits to the lowest-order Lifshits-Kosevich~\cite{schoenberg} expression for relative change in conductivity $\textit{f}_{\textrm{QO}}(B) \sim \Delta \sigma / \sigma_0$ as a function of temperature T and field B:
\begin{equation}
\textit{f}_{\textrm{QO}}(B) = R_D R_T \cos\left( 2\pi F / B\right)
\label{eq:lk}
\end{equation}
where $R_D = \exp(-\pi / \omega_c \tau_D)$, $\omega_c \equiv e B / m^*$ is the cyclotron frequency, $\tau_D$ is the Dingle lifetime, and $R_T = (2\pi^2 k_B T/\hbar \omega_c) / \sinh\left( 2\pi^2 k_B T/\hbar \omega_c \right)$.
We fit the entire data set using a single frequency $F$ = 255~T and lifetime $\tau_D$, plotted as the continuous curves in Fig.~\ref{f:invb}A.

\begin{table}[bt]
  \footnotesize
  \centering
  \caption{Properties of the \pccod\ film studied in this work, obtained by resistivity, Hall effect, and magnetoresistance (MR) measurements and analyses of magnetic quantum oscillations (lower entries).}
  	\begin{ruledtabular}
		\begin{tabular}{l c c l}
		Quantity	& Parameter	& Value	& Unit \\
		\hline
		Ce content & x & 0.14	& - \\
		Residual resistance ratio &	RRR & 10.2 & - \\
		Normal state resistivity (2 K) & $\rho_{xx}$ & 15 & $\mu \Omega$ cm \\
		Transition temperature & $T_c$ & 22 & K \\
		Hall coefficient (2 K) & $R_H$ & 0.9 & $10^{-9}$ $\Omega$ m/T \\
		Mobility from ... & & & \\
		- Hall effect (2 K) & $\mu_H$ & 0.0059 & T$^{-1}$ \\
		- Magnetoresistance ($<$50 K) & $\mu_{\textrm{MR}}$ & 0.013 & T$^{-1}$ \\
		- Dingle formula (2 K) & $\mu_D$ & 0.014 & T$^{-1}$ \\
		Dingle temperature (2 K) & $T_D$ & 44 & K \\
		QO frequency & $F$ & 255$\pm$10 & T \\
		QO effective mass & $m^*$ & 0.43$\pm$.05 & $m_e$ \\
		Fermi velocity & $v_F$ & 2.9 & $10^5$ m/s \\
	\end{tabular}
	\end{ruledtabular}
	\label{tab:params}
\end{table}

The large Dingle temperature ($T_D = \hbar / 2 \pi k_B \tau_{D} \approx $ 44~K) and short quasiparticle lifetime $\tau_{D} = 2.8 \times 10^{-14}$~s indicate that QO will only be visible on small FS pockets. The 255~T pocket cyclotron orbit size $\ell_c = 2 k_F/ \hbar e B$ is 12~nm at 100~T, while the mean free path $\ell_o = (\tau_D \hbar/ m^*) \sqrt{A_F/ \pi}$ is 8.2~nm. The cyclotron orbit size for an $\sim 11$~kT orbit, corresponding to either the low-temperature Hall effect in this PCCO film (see discussion below) or the large pockets predicted by band theory~\cite{massidda_electronic_1989,helm2009} would be a factor of 6-7 larger and not resolvable at 100~T in this film.

\begin{figure}[bt]
    \includegraphics[width=0.9\columnwidth]{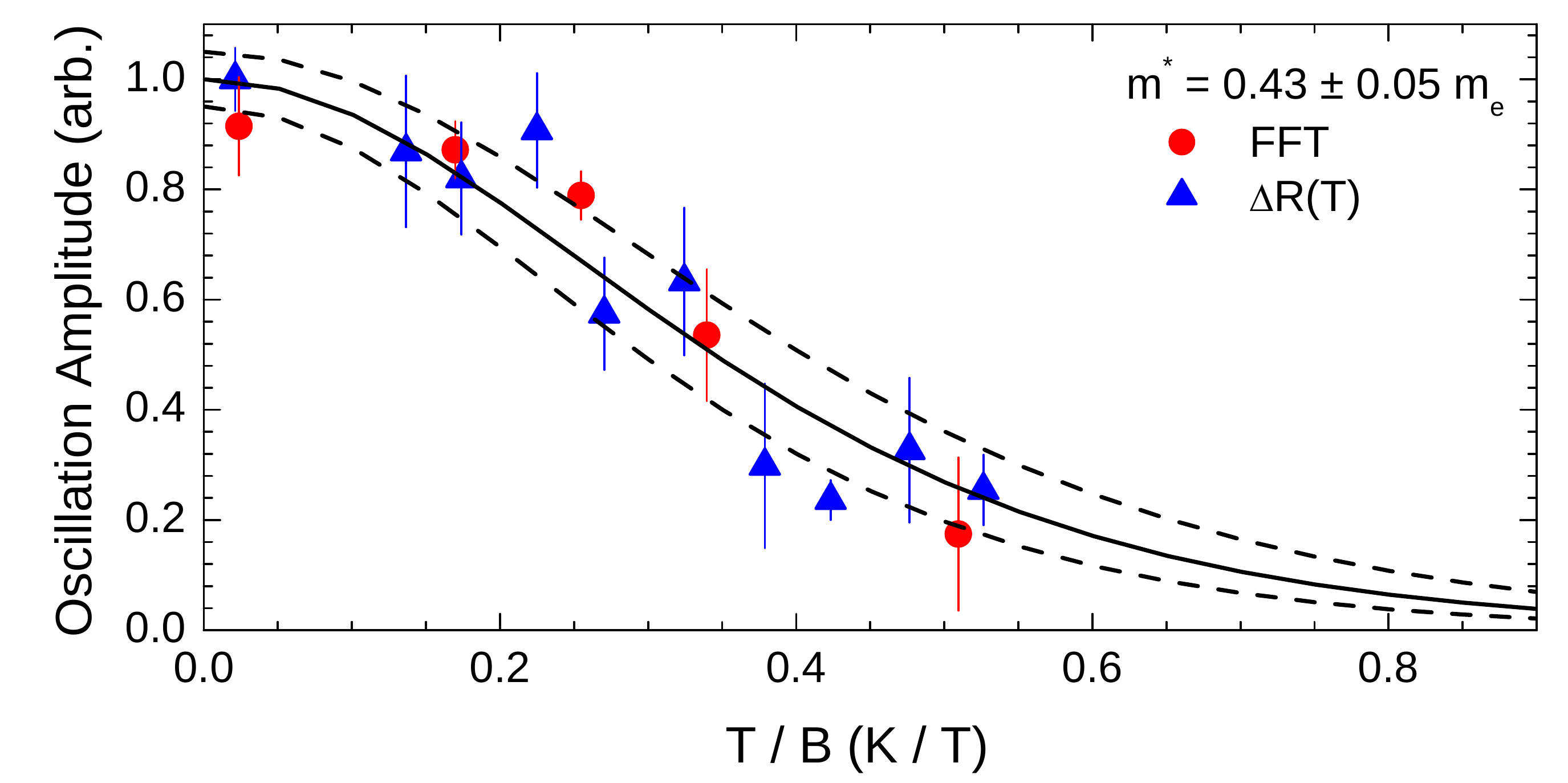}
\caption{(Color online) Decay of the quantum oscillation amplitude as a function of temperature, evaluated both using the FFT amplitude (circles) and resistance at fixed fields $\Delta R(T)$ (triangles); together they indicate a quasiparticle effective mass $m^* = 0.43 \pm 0.05 m_e$ (continuous and dashed lines).}
\label{f:fft}
\end{figure}

At fixed magnetic field the decrease in QO oscillation amplitude with increasing temperature is a direct measure of the quasiparticle effective mass $m^*$, a quantity that can be enhanced in proximity to a quantum critical point~\cite{ramshaw2015}. The QO amplitude is plotted versus temperature in Fig.~\ref{f:fft}, along with a fit yielding $m^* =  0.43 \pm 0.05$ m$_e$ consistent with both FFT spectrum amplitudes (red data) and analyses at fixed magnetic field (black data); also shown are $\pm1 \sigma$ error bars to the best fit (dashed lines). Here $F$ is comparable to that observed in NCCO with $x$ = 0.17 ($F_{\textrm{NCCO}} \approx 250$~T) in studies of bulk crystals~\cite{helm2015} that show evidence for a quantum critical point at a doping $x$ = 0.145. $R_H$ is comparable to that seen in $x$ = 0.17 PCCO films ($\approx 0.8 \times 10^{-9}$~$\Omega$-m/T) grown and oxygen reduced using other techniques~\cite{li2007, gauthier2007}. Differences in $m^*$ and $T_c$ that separate this film and PCCO and NCCO materials near $x$ = 0.17 will require further systematic study; given the typical decrease in charge carrier mobility with increasing $x$, we propose that films with fixed Ce concentration may, via suitable oxygen annealing, be used to continuously study the FS evolution towards overdoping. QCPs have been reported in PCCO near $x$ = 0.165 near the region of AFM and SC coexistence between $x$ = 0.12 and $x$ = 0.15~\cite{dagan2004, yu2007, charpentier2010}, as well as in La$_{2-x}$Ce$_x$CuO$_4$ through analysis of scaling phenomena~\cite{jin2011b, butch2012}, and it is likely that variation in nominal Ce concentation associated with this point in the phase diagram derives from differences in the synthesis and oxygen reduction processes.

\begin{figure}[tb]
    \includegraphics[width=1.0\columnwidth]{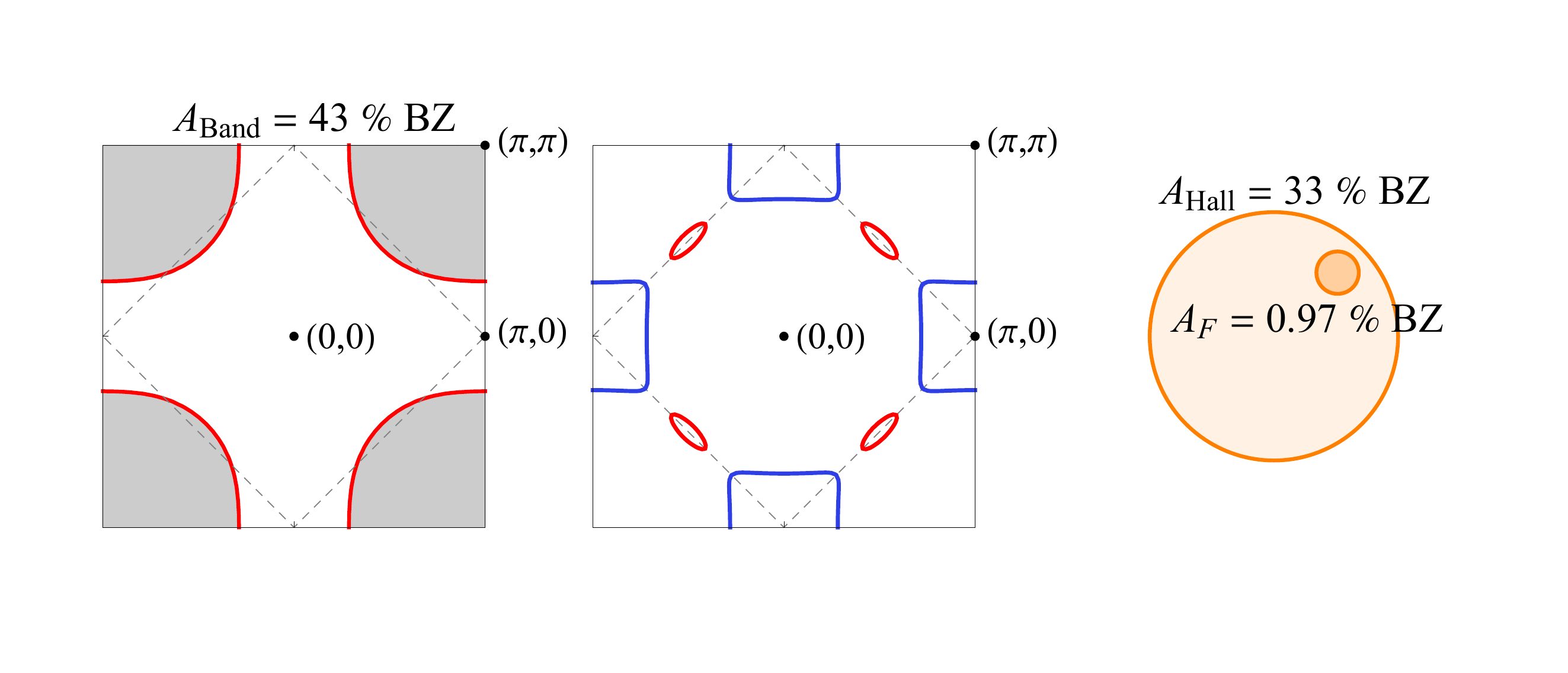}
\caption{(Color online) Schematic hole-like PCCO Fermi surface according to band theory for $x = 0.14$ (left), reconstruction picture at and below optimal doping (center), and the 2D FS areas (right) shown by magnetic quantum oscillation ($A_{F}$) and estimated using Hall effect measurements ($A_{\textrm{Hall}}$) as discussed in the text.}
\label{f:fs}
\end{figure}

The nature of the electronic order driving the reconstructed FS observed here remains to be settled. A comparison of prevailing FS schematics is shown in Fig.~\ref{f:fs}, including the large hole-like cylinder centered at ($\pi,\pi$) at left, and the reconstructed FS consistent with photoemission measurements and displaying two hole-like (red) and one electron-like (blue) pocket around the reduced (antiferromagnetic) Brillouin zone (dashed lines). As in NCCO~\cite{helm2009}, the QO frequency seen here is consistent with the hole pockets of the reconstructed FS, and not from the much larger electron pockets at ($\pi$, 0) (middle of Fig.~\ref{f:fs}). The absence of oscillations from the electron pockets remains a puzzle, especially in NCCO where a breakdown orbit along the full reconstructed FS is reported. The size of the observed pocket seen in these measurements is shown at right in Fig.~\ref{f:fs} (small region), to scale with the FS schematics. Also shown is an estimated FS area using the 2~K Hall effect according to Luttinger's theorm (large region). While the Hall effect is likely to be more complex in the presence of multiple bands within a reconstructed FS, ongoing studies of PCCO\cite{dagan2004, charpentier2010, lin_theory_2005} and hole-doped cuprates\cite{badoux2016} suggest a direct link between the evolution of $R_H$ and the FS.

The small FS pocket indicating a reconstruction in PCCO similar to that seen in NCCO, coupled with similar observations of Ce-free \pco\ films\cite{breznay_2015}, suggest that small $F\sim$250-350~T FS pockets are a universal feature of superconducting ``electron-doped'' materials. We cannot rule out the possibility of a field-induced FS reconstruction beginning at or below $\approx$50~T; although convincing evidence now exists for field-induced FS reconstruction arising from charge order in the hole-doped cuprates\cite{wu_magnetic_2011,gerber_three_2015,leboeuf_thermodynamic_2012}, the situation in the electron-doped materials\cite{dasilvaneto2015} remains less clear. The presence of a carrier density consistent with an unreconstructed FS pocket, along with a small pocket showing evidence for a reconstructed and multi-component FS, suggests that continued improvement in thin film materials quality may connect to the the overdoped, Fermi-liquid region of the phase diagram and provide direct insight into the fate of the electron ground state with doping.

\begin{acknowledgments}
We gratefully acknowledge the scientific and support staff of the Los Alamos National High Magnetic Lab Pulsed Field Facility for their technical assistance on this project, in particular the 100~T operations team. We also appreciate fruitful conversations with Toni Helm.
This work was supported by the Laboratory Directed Research and Development Program of Lawrence Berkeley National Laboratory under the U.S. Department of Energy Contract No. DE-AC02-05CH11231. A portion of this work was performed at the National High Magnetic Field Laboratory, which is supported by National Science Foundation Cooperative Agreement No. DMR-1157490 and the State of Florida. Portions of this work were also supported by the Gordon and Betty Moore Foundation's EPiQS Initiative through Grant GBMF4374, and the U.S. Department of Energy Office of Basic Energy Sciences ``Science at 100 T'' program.
\end{acknowledgments}

\bibliographystyle{unsrt}

\end{document}